\documentclass{aa}  

\usepackage{graphicx}
\usepackage{txfonts}
\begin{document} 


   \title{Inhomogeneity in the early Galactic chemical enrichment exposed by  beryllium abundances in extremely metal-poor stars\thanks{Based on observations collected at the European Organisation for Astronomical Research in the Southern Hemisphere under ESO programme 081.B-0151(A), 060.O-9025(B), 060.A-9022(A), 067.D-0439(A), 068.D-0094(A), 068.B-0475(A), 075.D-0048(A), 076.A-0463(A), and 0101.A-0229(A).}}

   \author{R. Smiljanic
          \inst{1}
          \and
          M.~G. Zych
          \inst{2}
          \and
          L. Pasquini
          \inst{3}
          }

   \institute{Nicolaus Copernicus Astronomical Center, Polish Academy of Sciences, ul. Bartycka 18, 00-716, Warsaw, Poland\\
              \email{rsmiljanic@camk.edu.pl}
         \and
         Faculty of Physics, University of Warsaw, ul. Pasteura 5, 02-093 Warsaw, Poland
         \and
         European Southern Observatory, Karl-Schwarzschild-Str. 2, 85748 Garching bei M\"unchen, Germany
             }

   \date{Received xx 2020; accepted xx 2020}

\titlerunning{Be in extremely metal-poor stars}
\authorrunning{Smiljanic, Zych \& Pasquini}

 
  \abstract
   {Abundances of beryllium in metal-poor stars scale linearly with metallicity down to [Fe/H] $\sim$ $-$3.0. In the stars where Be has been detected at this extremely metal-poor regime, an increased abundance scatter has been previously reported in the literature. This scatter could indicate a flattening of the relation between Be abundances and metallicity.}
   {Our aim is to perform a new investigation of Be abundances in extremely metal-poor stars and try to clarify whether a Be abundance plateau exists. We revisited the Be abundances in a sample of nine dwarfs with metallicities close to [Fe/H] $\sim$ $-$3.0. Additionally, we analysed the Be lines in the spectra of stars BPS BS 16968-0061 and CD-33 1173 for the first time.}
   {We took advantage of \emph{Gaia} DR2 parallaxes to refine values of the surface gravity of the stars. Updated values of surface gravity can have a significant impact on the determination of Be abundances. The other atmospheric parameters were computed using photometric and spectroscopic data. Abundances of Be were determined using spectrum synthesis and model atmospheres.}
   {Some of the stars indeed suggest a flattening. Over about a 0.5 dex range in metallicity, between [Fe/H] $\sim$ $-$2.70 and $-$3.26, the Be abundances stay mostly constant at about $\log(Be/H)$ $\sim$ $-$13.2 dex. Nevertheless, for several stars, we could only place upper limits that are below that level. Most of the sample stars are consistent with having been formed at the progenitor of the so-called \emph{Gaia}-Enceladus merger. Two out of the three stars likely formed in-situ are the ones that deviate the most from the linear relation.}
   {The mixed origin of these extremely metal-poor stars offers a clue to understanding the flattening. We suggest that our observations can be naturally understood as a consequence of the inhomogeneous star forming conditions in the early Galaxy. Without efficient mixing, the early interstellar medium would be characterised by a large scatter in Fe abundances at a given moment. Beryllium, on the other hand, because of its origins in cosmic-ray spallation, would have more homogeneous abundances (in a Galaxy-wide sense). We therefore suggest that the observed flattening of the Be-versus-metallicity relation reflects a stronger scatter in the Galactic Fe abundances at a given age.}

   \maketitle
%

\section{Introduction}

Beryllium is a light element with a single stable isotope, $^9$Be, which can only be produced by cosmic-ray spallation in the interstellar medium \citep{1970Natur.226..727R,1971A&A....15..337M}. Beryllium is produced in the break up of C, N, and O nuclei accelerated by supernovae (SNe) and/or SNe remnants \citep{2004A&A...424..747P,2018ARNPS..68..377T}. Standard primordial (or Big Bang) nucleosynthesis can produce only trace amounts of Be, at most at the level of $\log(Be/H)$\footnote{The notation indicates the relative abundance by number of Be atoms with respect to H atoms.} $\sim$ $-$17.6 dex \citep{2014JCAP...10..050C}.

Attempts to determine Be abundances in old metal-poor stars started with \citet{1984A&A...139..394M}, \citet{1988A&A...193..193R}, and \citet{1990ApJ...348L..57R,1992ApJ...388..184R}. It soon became apparent that the Be abundances scaled linearly with metallicity \citep{1992Natur.357..379G,1993AJ....106.2309B,1997A&A...319..593M}. The slope of this relation has been shown to be very close to one \citep{2009ApJ...701.1519R,2009A&A...499..103S,2009MNRAS.392..205T,2011ApJ...743..140B}. Galactic chemical evolution models can successfully explain the linear relation if the chemical composition of the accelerated cosmic-ray material is adequate \citep{2012A&A...542A..67P}.

However, the existence of the linear relation at the extremely metal-poor regime (around and below [Fe/H]\footnote{The bracket notation indicates the ratio of the abundances by number in a star with respect to the same ratio in the Sun: [A/B] = $\log$ [N(A)/N(B)]$_{\star}$ - $\log$ [N(A)/N(B)]$_{\odot}$.} $\sim$ $-$3.0) has been subject to some discussion. \citet{2000A&A...364L..42P,2000A&A...362..666P} were the first to explore this metallicity regime and detected Be in two stars, LP 815-43 and G 64-12. An upper limit of the Be abundance was also determined for CD$-$24 17504. These authors suggested a possible deviation from the linear relation below [Fe/H] $\sim$ $-$3.0. \citet{2006ApJ...641.1122B} added another star in this regime, G64-37, and argued the evidence showed increased dispersion in the Be abundances but not a flattening of the relation. \citet{2009ApJ...701.1519R} presented Be abundances for 19 stars with [Fe/H] $<$ $-$2.5 (five stars with [Fe/H] $<$ $-$ 3.0) and argued there could be a change of the slope between Be and Fe at the lowest metallicities.

The evidence supporting the flattening of the relation was made weaker by the upper limit of the Be abundance in BD+44 493, a star with [Fe/H] = $-$3.7, derived by \citet{2009ApJ...698L..37I,2013ApJ...773...33I} and later refined by \citet{2014ApJ...790...34P}. Similar evidence against the flattening is given by the more recent Be upper limit for star 2MASS J1808-5104 \citep{2019A&A...624A..44S}. In both cases, the inferred limits are below the level of the possible plateau.

The interest in understanding whether a Be plateau exists also comes from the possibility of it having a primordial origin. For example, \citet{1997ApJ...488..515O} and \citet{2001PhRvD..64b3510J} show that inhomogeneous primordial nucleosynthesis could result in significant Be production, although still an order of magnitude below the observed Be detections at [Fe/H] $\sim$ $-$3.0. Another way to boost production of primordial Be would be with the existence of a long-lived, negatively charged massive particle \citep{2007arXiv0712.0647P,2008JCAP...11..020P,2014ApJS..214....5K,2017IJMPE..2641004K}. Other scenarios (not connected to primordial production) that can result in a Be plateau have also been proposed. For example, high-Be abundances could result from the effects of pre-Galactic cosmic rays in the intergalactic medium \citep{2008ApJ...673..676R,2008ApJ...681...18K} or from very massive stars in the early Galaxy \citep{1998A&A...337..714V}.

The interest in probing the possible flattening of the Be abundances at [Fe/H] $\sim$ $-$3.0 has motivated us to identify additional stars to be analysed at this regime. Thus, we have obtained new near-ultraviolet spectra for BPS BS 16968-061, with [Fe/H] = $-$3.05 \citep{2007A&A...462..851B} and analyse its Be lines for the first time. We also present the Be abundance for star CD-33 1173 for the first time. In the latter case, we made use of archival spectra not obtained by us. In addition, motivated by the availability of parallaxes from the data release 2 (DR2) of \emph{Gaia} \citep{2016A&A...595A...1G,2018A&A...616A...1G}, we decided to revisit the Be abundances in many of the previously analysed extremely metal-poor stars. With \emph{Gaia} parallaxes, we can improve the determination of the surface gravity of these stars, the parameter that is most important for the determination of Be abundances from the ionised \ion{Be}{ii} spectral lines.

This paper is organised as follows. The observational spectroscopic, photometric, and astrometric data are described in Section \ref{sec:data}. The determination of atmospheric parameters and abundances is described in Section \ref{sec:analysis}. In Section \ref{sec:discussion}, we discuss the results, and our conclusions are summarised in Section \ref{sec:end}.

\section{Observational data, kinematics, and orbits}\label{sec:data}

All spectra discussed here were obtained using the Ultraviolet and Visual Echelle Spectrograph \citep[UVES,][]{2000SPIE.4008..534D}, fed by the Unit Telescope 2 of the Very Large Telescope (VLT) of the European Southern Observatory (ESO) at Cerro Paranal, Chile. The blue spectra used in our analysis were obtained with the standard setup centred at 346 nm, resulting in a wavelength coverage between 303 and 388 nm. Data reduction was performed using the UVES pipeline \citep{2000Msngr.101...31B} within the ESO Reflex environment \citep{2013A&A...559A..96F}.

Star BPS BS 16968-0061, whose Be abundance is investigated here for the first time, was observed 12 times, for a total of 20 hours, between April and August of 2008. The log book of these observations is given in Table \ref{tab:bs16} together with the measured radial velocities (RVs). The mean heliocentric RV of the star was found to be $-$80.8 $\pm$ 0.9 km s$^{-1}$, which seems consistent with no RV variation. We note here that \citet{2014ApJ...788..180C} lists RV = $-$80.7 km s$^{-1}$ for this star, in excellent agreement with our value. The spectra of all other sample stars were obtained from the ESO data archive.

The sample is identified in Table \ref{tab:sample} where identification numbers from 2MASS \citep[Two Micron All Sky Survey,][]{2006AJ....131.1163S} and \emph{Gaia} DR2 are also given. For this sample, the parallax errors are small and it should be possible to compute distances by a simple inversion of the parallax \citep{2018A&A...616A...9L}. Nevertheless, we decided to adopt the estimates from the Bayesian analysis of \citet{2018AJ....156...58B}, which are given in Table \ref{tab:kinem}. The uncertainty on the value of the distance calculated by that method is not symmetric; the lower or upper bounds are different. For the purpose of propagating the uncertainties, we adopted the larger of the two bounds as the typical error value. 

These distances were used to estimate colour excess values, E($B-V$), through the STructuring by Inversion the Local Interstellar Medium (STILISM) database\footnote{\url{https://stilism.obspm.fr/}} \citep{2014A&A...561A..91L,2017A&A...606A..65C}. All stars are within the limits of the extinction maps, except for BS 16968-061. For BS 16968-061, we adopted the value determined by \citet{2009A&A...497..497G}. The E($B-V$) values are given in Table \ref{tab:kinem}.

Parallaxes and proper motions from \emph{Gaia}, together with RVs measured from our spectra, were used to compute the heliocentric Galactic space-velocity components ($U$, $V$, and $W$) following \citet{1987AJ.....93..864J}. The components are in the right-hand system: $U$ is positive towards the Galactic centre, $V$ is positive towards the Galactic rotation, and $W$ is positive towards the Galactic north pole. For the circular speed of the Sun and the solar motion relative to the local standard of rest (LSR), we adopted the values from \citet{2017MNRAS.465...76M}: $v_0$ = 232.8 km s$^{-1}$ and ($U$, $V$, $W$)$_{\rm LSR}$ = (8.6 $\pm$ 0.9, 13.9 $\pm$ 1.0, 7.1 $\pm$ 1.0) km s$^{-1}$. The velocities are given in Table \ref{tab:kinem}.

In addition, we integrated the orbits of the stars in the Galaxy using code {\sf GalPot}\footnote{\url{https://github.com/PaulMcMillan-Astro/GalPot}} of \citet{2017MNRAS.465...76M}, which is based on the code by \citet{1998MNRAS.294..429D}. For the calculations, we adopted the the best-fit Galactic model found by \citet{2017MNRAS.465...76M}. A total of 1000 Monte Carlo simulations were run to estimate the errors in the orbital parameters. The values of all orbital quantities used here are listed in Table \ref{tab:orbits}.

\begin{table}
\caption{Log book of the observations of BPS BS 16968-0061 and the measured radial velocities.}             
\label{tab:bs16}      
\centering\small      
\begin{tabular}{cccc}     
\hline\hline       
Obs. Date. & UTC at start & Exposure & Heliocentric \\
                  &                     &   time &  RV \\
(yyyy-mm-dd) & (hh:mm:ss.ss) & (s) & (km s$^{-1}$) \\
\hline                    
2008-04-19 & 06:48:45.23 & 5400 & $-$80.4 \\
2008-04-20 & 03:56:24.89 & 5400 & $-$81.2 \\
2008-04-23 & 05:35:10.44 & 5400 & $-$82.9 \\
2008-04-26 & 05:16:19.88 & 5400 & $-$80.3 \\
2008-04-27 & 03:09:31.00 & 5400 & $-$80.0 \\
2008-04-27 & 04:44:36.39 & 5400 & $-$80.5 \\
2008-05-13 & 02:34:00.00 & 5400 & $-$81.1 \\
2008-05-13 & 04:09:57.00 & 5400 & $-$81.0 \\
2008-05-13 & 05:43:22.00 & 5400 & $-$81.5 \\
2008-05-26 & 02:56:04.00 & 5400 & $-$81.2 \\
2008-07-19 & 00:42:13.00 & 5400 & $-$79.7 \\
2008-08-01 & 23:46:29.00 & 5400 & $-$80.1 \\
\hline                  
\end{tabular}
\end{table}
\begin{table*}
        \caption{Sample stars: identifications, coordinates, and visual magnitude.}             
        \label{tab:sample}      
        \centering\small          
        \begin{tabular}{lccccc}     
                \hline\hline       
            Star & RA         & Dec          & 2MASS ID           & Gaia                         &   $V$                  \\
& (deg)      & (deg)        &                    &  source name                 &    (mag)               \\
\hline                    
LP831-70         & 046.522670 & $-$22.321650 & 03060544$-$2219179 & Gaia DR2 5078491312157905280 & 11.608$\pm$0.021$^1$ \\
CD-33 1173       & 049.897126 & $-$32.845367 & 03193531$-$3250433 & Gaia DR2 5054340333095950464 & 10.904$\pm$0.021$^1$ \\
BD+03 740        & 075.319296 & +04.110301   & 05011663+0406370   & Gaia DR2 3238602707816488064 & 9.808$^2$            \\
BD+24 1676       & 112.671913 & +24.086176   & 07304125+2405102   & Gaia DR2 866863321051682176  & 10.743$\pm$0.027$^1$ \\
BD+09 2190       & 142.314886 & +08.633416   & 09291557+0838002   & Gaia DR2 588856788129452160  & 11.144$\pm$0.023$^1$ \\
BD-13 3442       & 176.711034 & $-$14.111995 & 11465064$-$1406431 & Gaia DR2 3572742676589941760 & 10.274$\pm$0.060$^1$ \\
G64-12           & 205.010399 & $-$00.038567 & 13400249$-$0002188 & Gaia DR2 3662741860852094208 & 11.457$\pm$0.008$^1$ \\
G64-37           & 210.625364 & $-$05.651358 & 14023008$-$0539048 & Gaia DR2 3643857920443831168 & 11.125$\pm$0.015$^1$ \\
BPS BS 16968-061 & 225.275458 & +03.712643   & 15010610+0342455   & Gaia DR2 1154852693702722432 & 13.219$\pm$0.024$^1$   \\
LP815-43         & 309.555447 & $-$20.436260 & 20381330$-$2026105 & Gaia DR2 6856600660237061504 & 10.869$\pm$0.018$^1$ \\
CD-24 17504      & 346.834295 & $-$23.876556 & 23072023$-$2352356 & Gaia DR2 2383484851010749568 & 12.062$\pm$0.041$^1$ \\
                \hline                  
        \end{tabular}
        \tablefoot{Coordinates are at epoch J2000 from the 2MASS catalogue. The $V$ magnitudes are from the following: (1) \citet{2015AAS...22533616H} and (2) \citet{1987A&AS...71..413M}.}
\end{table*}
\begin{table*}
\caption{Distances, colour excess, and Galactic space velocities.}             
\label{tab:kinem}      
\centering\small          
\begin{tabular}{lrcrrr}     
\hline\hline       
Star & \multicolumn{1}{c}{Distance}  & E(B-V) & \multicolumn{1}{c}{U} &  \multicolumn{1}{c}{V} &  \multicolumn{1}{c}{W} \\
 & \multicolumn{1}{c}{(pc)} & (mag) &  \multicolumn{1}{c}{(km s$^{-1}$)} & \multicolumn{1}{c}{(km s$^{-1}$)}  & \multicolumn{1}{c}{(km s$^{-1}$)}  \\
\hline                    
LP831-70         & 241.1$\pm$2.4  & 0.011$\pm$0.016 &     55.9$\pm$0.5 &     80.2$\pm$1.8  &     90.0$\pm$1.0 \\
CD-33 1173       & 217.4$\pm$1.3  & 0.005$\pm$0.014 &    136.4$\pm$0.9 &     77.4$\pm$1.0  &  $-$17.5$\pm$0.9 \\ 
BD+03 740        & 170.8$\pm$2.1  & 0.019$\pm$0.019 & $-$118.7$\pm$1.0 &     37.6$\pm$2.1  &  $-$17.3$\pm$0.7 \\
BD+24 1676       & 256.2$\pm$3.2  & 0.013$\pm$0.018 &    335.7$\pm$1.7 &  $-$14.1$\pm$4.0  &      2.6$\pm$1.0 \\
BD+09 2190       & 269.9$\pm$3.3  & 0.021$\pm$0.017 &    202.2$\pm$4.2 & $-$217.1$\pm$3.9  &    202.4$\pm$0.7 \\
BD-13 3442       & 213.2$\pm$4.1  & 0.026$\pm$0.016 &    138.7$\pm$2.2 &     58.9$\pm$2.2  &  $-$32.1$\pm$2.5 \\
G64-12           & 264.0$\pm$6.2  & 0.025$\pm$0.021 &     22.7$\pm$4.1 & $-$119.2$\pm$5.9  &    400.6$\pm$0.9 \\
G64-37           & 236.0$\pm$5.5  & 0.025$\pm$0.021 &    200.4$\pm$3.5 & $-$135.2$\pm$8.4  & $-$148.6$\pm$5.2 \\
BPS BS 16968-061 & 959.0$\pm$39.2 & 0.05            &  $-$52.1$\pm$0.8 &  $-$50.3$\pm$12.1 &  $-$38.7$\pm$1.0 \\
LP815-43         & 242.2$\pm$2.9  & 0.028$\pm$0.022 &    139.0$\pm$1.8 &   $-$2.6$\pm$3.0  &     33.9$\pm$0.6 \\
CD-24 17504      & 312.5$\pm$4.9  & 0.012$\pm$0.016 &  $-$99.4$\pm$2.4 &  $-$75.2$\pm$5.6  & $-$262.3$\pm$2.4 \\
\hline                  
\end{tabular}
\end{table*}

\begin{table*}
\caption{Orbital parameters.}             
\label{tab:orbits}      
\centering\small          
\begin{tabular}{lrrrrrr}     
\hline\hline       
Star & Rmin & Rmax & Zmax & Eccen. & Orbital energy & L$_z$\\
        & (kpc) & (kpc) & (kpc) &  & (10$^5$ km$^2$ s$^2$) & (km s$^{-1}$ kpc)  \\
\hline                    
LP831-70 & 1.85$\pm$0.05 & 8.70$\pm$0.01 & 3.45$\pm$0.06 & 0.64$\pm$0.01 & -1.73482$\pm$1.40E-3 & -650$\pm$16  \\
CD-33 1173 & 1.44$\pm$0.02 & 10.10$\pm$0.02 & 0.63$\pm$0.01 & 0.75$\pm$0.01 & -1.69912$\pm$0.84E-3 & -617$\pm$9 \\ 
BD+03 740 & 0.66$\pm$0.04 & 9.51$\pm$0.02 & 0.34$_{-0.06}^{+0.42}$ & 0.87$\pm$0.01 & -1.74594$\pm$0.153E-3 & -309$\pm$17\\
BD+24 1676 & 0.26$\pm$0.06 & 25.76$\pm$0.31 & 0.31$_{-0.04}^{+0.92}$ & 0.98$\pm$0.01 & -1.24180$\pm$6.27E-3 & 155$\pm$34 \\
BD+09 2190 & 4.81$\pm$0.03 & 27.3$\pm$1.2 & 17.1$\pm$0.6 & 0.63$\pm$0.01 & -1.16878$\pm$1.670E-2 & 1853$\pm$32 \\
BD-13 3442 & 1.01$\pm$0.05 & 10.0$\pm$0.06 & 0.80$_{-0.04}^{+0.10}$ & 0.82$\pm$0.01 & -1.71142$\pm$2.99E-3 & -452$\pm$20 \\
G64-12 & 2.43$\pm$0.11 & 43.8$\pm$0.9 & 42.0$\pm$0.6 & 0.69$\pm$0.01 & -0.96163$\pm$8.13E-3 & 975$\pm$43 \\
G64-37 & 3.01$_{-0.19}^{+0.15}$ & 15.9$_{-0.8}^{+0.9}$ & 8.2$_{-0.4}^{+2.3}$ & 0.64$\pm$0.01 & -1.43028$\pm$2.55E-2 &  1116$\pm$67  \\
BPS BS 16968-061 & 0.95$\pm$0.26 & 7.86$^{+0.09}_{-0.04}$ & 1.20$^{+1.98}_{-0.07}$ & 0.78$\pm$0.06 & -1.82872$\pm$5.76E-3 & 400$\pm$97\\
LP815-43 & 0.04$\pm$0.04 & 9.72$\pm$0.06 & 4.62$_{-0.15}^{+0.09}$ & 0.99$\pm$0.01 & -1.73759$\pm$2.58E-3 &    22$\pm$28 \\
CD-24 17504 & 2.16$\pm$0.15 & 14.0$\pm$0.5 & 13.5$\pm$0.4 & 0.73$\pm$0.01 & -1.41673$\pm$1.212E-2 & 630$\pm$47 \\
\hline                  
\end{tabular}
\tablefoot{Uncertainties come from Monte Carlo simulations that do not necessarily return symmetric distributions. Values from the 15.9 and 84.1 percentiles are reported. For the first three quantities, if the uncertainties are within 0.02 kpc, we report the bigger one as the typical value. When these values are very different, both are reported. If the percentiles are below 0.01 kpc, we assumed the minimum uncertainty to have a value of 0.01 kpc. L$_z$ stands for the orbital angular momentum component, perpendicular to the orbit plane.}
\end{table*}
\begin{table*}
\caption{Atmospheric parameters and abundances (with uncertainties).}             
\label{tab:atm}      
\centering\small          
\begin{tabular}{lcccccc}     
\hline\hline       
Star & $T_{\rm eff}$ & $\log~g$ & [Fe/H] & $\log({\rm Be/H})_{\rm 3130}$ & $\log({\rm Be/H})_{\rm 3131}$ & S/N \\ 
 & (K) & (dex) & (dex)  & (dex) &  (dex) & (per px) \\ 
\hline                    
LP831-70 & 6350$\pm$125 & 4.42$\pm$0.05 & $-$3.00$\pm$0.08 & $\leq$ $-$13.3 & $\leq$ $-$13.4 & 90 \\%
CD-33 1173 & 6605$\pm$115 & 4.30$\pm$0.04 & $-$2.95$\pm$0.07 & $\leq$ $-$13.8 & $-$13.2 & 200 \\%
BD+03 740 & 6400$\pm$150 & 4.00$\pm$0.05 & $-$2.72$\pm$0.08 & $\leq$ $-$13.5 & $-$13.4 & 180 \\ %
BD+24 1676 & 6500$\pm$120 & 4.05$\pm$0.04 & $-$2.35$\pm$0.06 & $-$12.9 & $-$12.8* & 80 \\ %
BD+09 2190 & 6480$\pm$130 & 4.16$\pm$0.05 & $-$2.70$\pm$0.08 & $-$13.2 & $\leq$ $-$13.3 & 170 \\ %
BD-13 3442 & 6505$\pm$155 & 4.00$\pm$0.05 & $-$2.88$\pm$0.05 & $-$13.2 & $-$12.9* & 150 \\ 
G64-12 & 6490$\pm$145 & 4.30$\pm$0.05 & $-$3.26$\pm$0.03 & $-$13.2* & $-$13.1 & 120 \\ 
G64-37 & 6630$\pm$160 & 4.30$\pm$0.06 & $-$3.11$\pm$0.07 & $-$13.1 & $-$13.2* & 190 \\ 
BPS BS 16968-061 & 6350$\pm$155 & 3.80$\pm$0.06 & $-$2.98$\pm$0.10 & $\leq$ $-$13.8  & $-$13.2* & 200 \\ %
LP815-43 & 6520$\pm$130 & 4.15$\pm$0.05 & $-$2.78$\pm$0.06 & $\leq$ $-$13.4* & $\leq$ $-$13.4* & 110 \\ %
CD-24 17504 & 6390$\pm$135 & 4.38$\pm$0.05 & $-$3.26$\pm$0.10 & $\leq$ $-$13.5 & $\leq$ $-$13.0 & 60 \\ %
\hline                  
\end{tabular}
\tablefoot{Values marked with an asterisk (*) should be considered more uncertain, as the observed line profiles disagree with the synthetic ones (the profiles are strongly affected by noise, or either broader or narrower than expected). S/N is the signal-to-noise ratio per pixel at 3133 \AA.}
\end{table*}

\begin{table*}
\caption{Equivalent widths of \ion{Fe}{ii} lines in m\AA.}             
\label{tab:eqw}      
\centering\small          
\begin{tabular}{lccccccc}     
\hline\hline       
Star & 4178.86 & 4233.17 & 4583.84 & 4923.93 & 5018.44 & 5169.03 & 5316.62 \\
\hline                    
LP831-70 & -- & 10.7 & -- & 19.7 & 25.6 & 31.4 & 5.0 \\
CD-33 1173 & 2.8 & 12.0 & -- & 22.5 & 28.3 & 33.5 & 5.5 \\
BD+03 740 & -- & -- & -- & 38.3 & 45.5 & 53.1 & 12.5 \\
BD+24 1676 & 12.6 & 34.6 & -- & 51.2 & 59.2 & 68.4 & 20.1 \\
BD+09 2190 & -- & -- & -- & 35.5 & 43.6 & 50.8 & 11.6 \\
BD-13 3442 & -- & -- & -- & 36.8 & 44.4 & 51.6 & 10.7 \\
G64-12 & -- & -- & -- & 13.9 & 18.2 & 24.5 & -- \\
G64-37 & -- & 8.8 & -- & 16.6 & 22.8 & 27.8 & -- \\
BPS BS 16968-061 & 5.8 & 16.0 & -- & 30.0 & 36.4 & 42.1 & 9.8 \\
LP815-43 & -- & -- & -- & 30.2 & 39.2 & 46.2 & 9.0 \\
CD-24 17504 & -- & 5.8 & 5.3 & 12.0 & 15.7 & 21.0 & -- \\
\hline                  
\end{tabular}
\end{table*}

\section{Spectral analysis}\label{sec:analysis}

\subsection{Atmospheric parameters}

Effective temperatures ($T_{\rm eff}$) were determined using the infrared flux method (IRFM) calibrations derived by \citet{2009A&A...497..497G} based on the 2MASS $J$, $H$, and $Ks$ magnitudes and the $V$ magnitudes listed in Table \ref{tab:sample}. As the calibrations are metallicity dependent, the procedure was necessarily iterative. A first guess of $T_{\rm eff}$ was computed assuming [Fe/H] = $-$3.0 and revised after the metallicity was determined as described below. This procedure was repeated until convergence, which usually took two or three iterations.

To compute the surface gravity, we made use of the equation
{
\begin{equation}
\log(\frac{g_{\star}}{g_{\odot}}) = \log(\frac{M_{\star}}{M_{\odot}}) + 4\times\log(\frac{T_{\rm eff \star}}{T_{\rm  eff  \odot}}) - \log\, (\frac{L_{\star}}{L_{\odot}}),
\end{equation}
}
\noindent where we assumed the sample stars to have 0.8 M$_{\odot}$ and used the solar values ($T_{\rm eff}$, $\log~g$) = (5777 K, 4.44). The stellar luminosities were computed assuming the bolometric magnitude of the Sun to be 4.75 mag. Bolometric magnitudes for the sample stars were estimated using the 2MASS $Ks$ magnitudes, the distances based on \emph{Gaia} parallaxes, bolometric corrections computed with the tools provided by \citet{2014MNRAS.444..392C,2018MNRAS.475.5023C}, and reddening computed from the colour excess of Table \ref{tab:kinem} using the relations of \citet{2004AJ....128.2144M}. As the $\log~g$ values also depend on $T_{\rm eff}$, the determination of the surface gravity was also iterative.

Metallicities were determined using equivalent widths of up to seven \ion{Fe}{ii} lines. Abundances were computed under the local thermodynamical equilibrium (LTE) approximation and using the MARCS one-dimensional model atmospheres \citep{2008A&A...486..951G} with the {\sf Turbospectrum} code \citep{2012ascl.soft05004P}. We chose to rely on \ion{Fe}{ii} lines as they are formed on deep atmospheric layers, and the impact of granulation effects, as computed in three-dimensional model atmospheres, is minor \citep{2001A&A...372..601A}. Microturbulence was assumed to be 1.5 km s$^{-1}$. The final atmospheric parameters are listed in Table \ref{tab:atm}.

The \ion{Fe}{ii} equivalent widths were measured using the {\sf splot} task with {\sf IRAF}\footnote{IRAF is distributed by the National Optical Astronomy Observatory, which is operated by the Association of Universities for Research in Astronomy (AURA) under a cooperative agreement with the National Science Foundation: \url{http://ast.noao.edu/data/software}.} and are listed in Table \ref{tab:eqw}. The oscillator strengths of the \ion{Fe}{ii} lines are from \citet{2009A&A...497..611M}, and the broadening by neutral hydrogen uses data from the calculations of \citet{2005A&A...435..373B}. The remaining atomic data are from the Kurucz database \citep{K13} and were obtained through VALD\footnote{\url{http://vald.astro.uu.se/}} \citep[Vienna Atomic Line Database,][]{1995A&AS..112..525P,2015PhyS...90e4005R}.

The errors of the metallicities are the standard deviation of the individual line abundances. For $T_{\rm eff}$ and $\log~g$, the errors come from 1000 Monte Carlo runs assuming Gaussian uncertainties in the observed magnitudes, colour excess (with eventual negative values turned into null values), metallicities, parallaxes, and bolometric corrections. For the bolometric corrections, which themselves depend on the atmospheric parameters, we assumed a typical uncertainty of $\pm$0.06 mag. This value was itself derived from 1000 Monte Carlo runs with typical uncertainties of $\pm$120 K in $T_{\rm eff}$, $\pm$0.05 dex in $\log~g$, and $\pm$0.08 in [Fe/H].

\subsection{Beryllium abundances}

The Be abundances were derived using spectrum synthesis, also with {\sf Turbospectrum} and MARCS models. For the \ion{Be}{ii} lines, the energy levels and central wavelengths are from the critical compilation by \citet{2005PhyS...72..309K}, oscillator strengths are from \citet{1977A&AS...27..489B}, and the broadening by neutral hydrogen uses data from the calculations of \citet{2000A&AS..142..467B}. The list of OH lines came from the Kurucz database \citep{KOH}. We adopted the $^{12}$CH line list of \citet{2014A&A...571A..47M}. Additional atomic data were obtained from the VALD database in the range between 3120 and 3145 \AA. The strong \ion{V}{ii} (at 3130.27 \AA) and \ion{Ti}{ii} (at 3130.80 \AA) lines, which are found around the Be lines, have oscillator strengths from \citet{2014ApJS..214...18W} and \citet{2001ApJS..132..403P}, respectively, with broadening constants from the Kurucz database \citep{K10}.

As the Be lines in the spectra of our stars are very weak, the proper fitting of the lines was challenging. Our procedure was as follows. The first step was quantifying the total broadening needed to fit the spectral region of the Be lines. That was done with the help of the strong nearby \ion{Ti}{ii} and \ion{V}{ii} lines and the OH feature at 3130.57 \AA. The abundances of Ti, V, and O, and the full width at half maximum for a Gaussian broadening were simultaneously varied until a reasonable fit to these three strong lines was achieved. This first fit of the region fixed a value for the broadening.

The second step started with the computation of 13 synthetic spectra, for each star, with [Be/Fe] values between $-$0.6 and +0.6 in steps of 0.1 dex, and one spectrum computed without Be. The spectra were computed with a sampling of 0.02 \AA\ and broadened to the value that was determined before. The observed spectra were resampled to the same dispersion solution as the synthetic spectra. We then searched for the synthetic spectrum that minimised the relative variance in a region of $\pm$0.08 \AA\ around the centre of each Be line. The fitting procedure was conducted separately for each of the two Be lines.

After the first variance minimum was found, 18 additional spectra were computed to cover all values between $\pm$0.10 dex from the best abundance, in steps of 0.01 dex. The variance minimisation was then repeated to confirm the best [Be/Fe] value. This best abundance value was, nevertheless, still rounded to the first decimal digit. This was done to take into account that for such low Be abundances, variations of $\pm$0.01 can not be properly distinguished. 

The next step was to decide whether each Be line was appropriately detected or an upper limit should be reported. To do that, we analysed the distribution of normalised residuals of the observed spectrum with respect to both the best-fit synthetic spectrum and the spectrum computed without beryllium. The residuals were normalised using the error expected because of the signal-to-noise ratio (S/N).

We first performed a two-sample Kolmogorov-Smirnov test, on the two distributions of residuals. The null hypothesis was that there is no difference between the two distributions. If the null hypothesis could not be rejected, we decided the line could not be detected and reported an upper limit. Otherwise, we had a first indication that a detection was possible. To confirm the detection, we ran a one-sample Kolmogorov-Smirnov test using the normalised residuals of the best fit and comparing it to the cumulative distribution function of a Gaussian that has mean equal to zero and variance equal to one. A good fit to the spectral line should provide normalised residuals with that normal distribution. If instead the null hypothesis that there is no difference between the two distributions were rejected, we assumed that the line could not be well fit and again reported an upper limit. Conservatively, because the lines are weak and variations are too small, the upper limit was taken to be 0.1 dex above the best [Be/Fe] found in the fitting procedure.

The choice of best-fit Be abundance required revision in a few cases. For star CD-24 17504, only upper limits could be determined because of the low S/N. The line at 3130 \AA\ seems absent. In this case, several synthetic spectra with low but different [Be/Fe] values result in similar fits with respect to the synthetic spectrum without Be. We decided to set the upper limit at the abundance value that would ensure the line be visually detected, if present.

\begin{figure*}[th]
        \centering
        \includegraphics[height = 9cm]{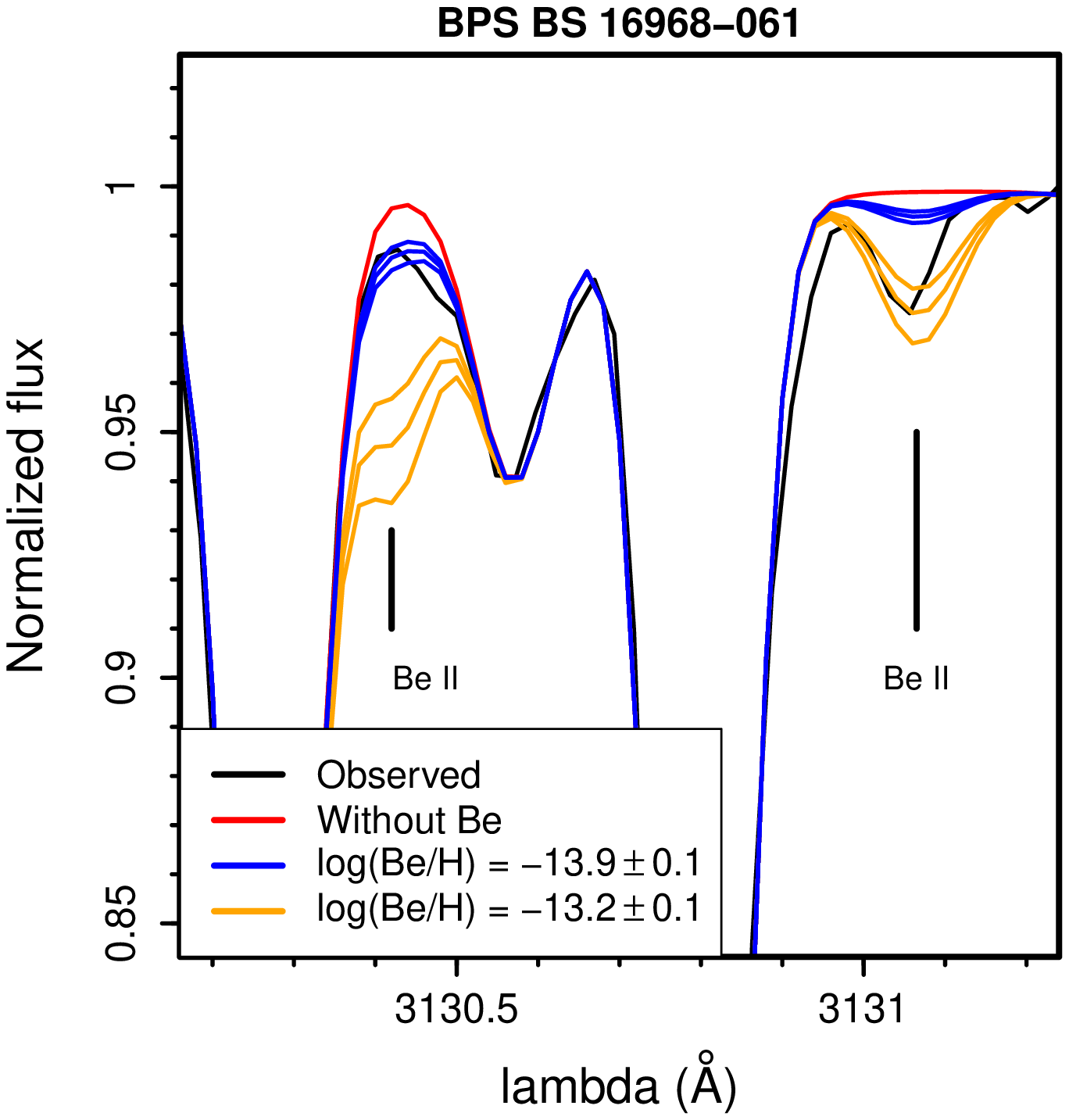}
        \includegraphics[height = 9cm]{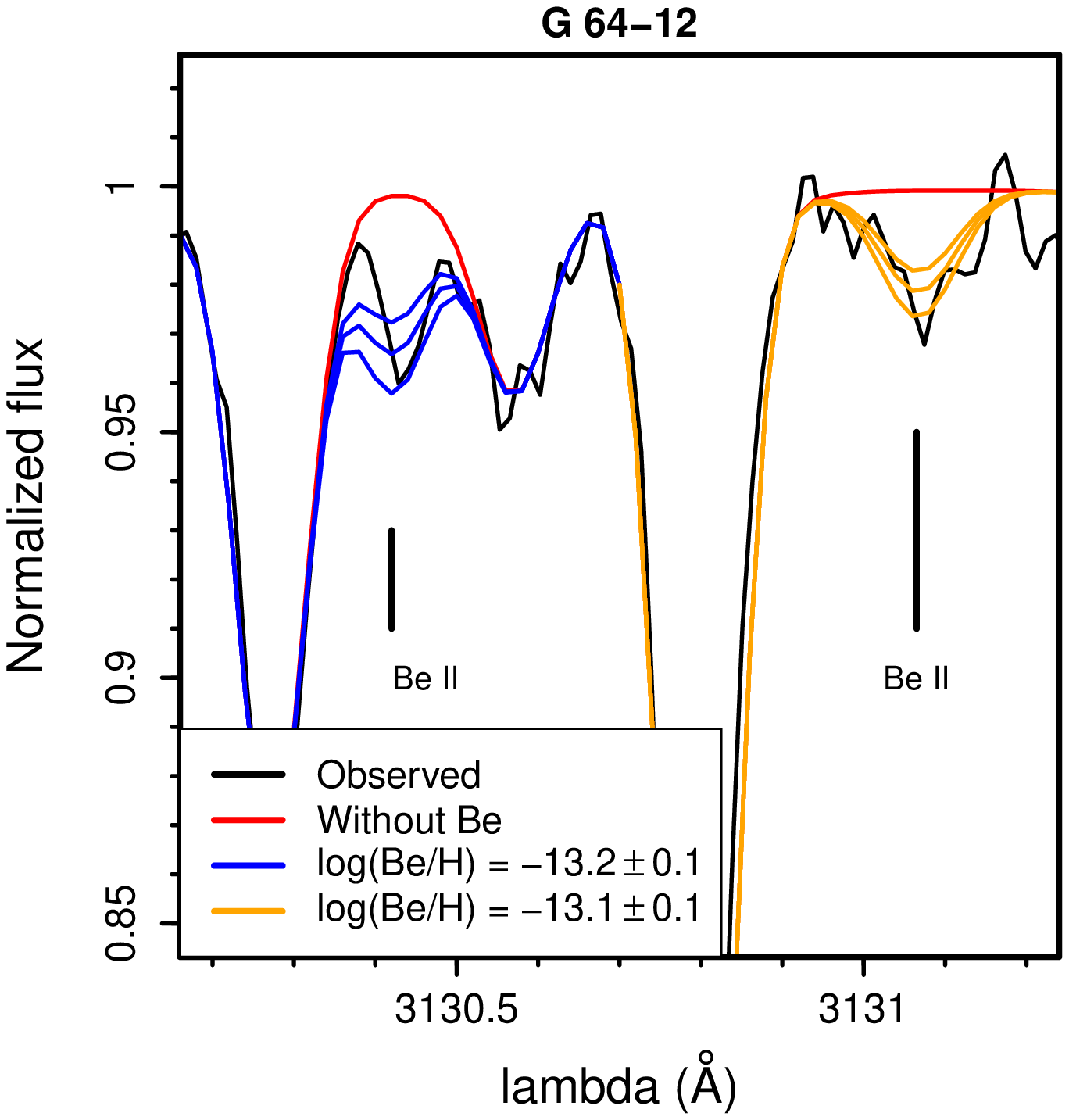}
        \includegraphics[height = 9cm]{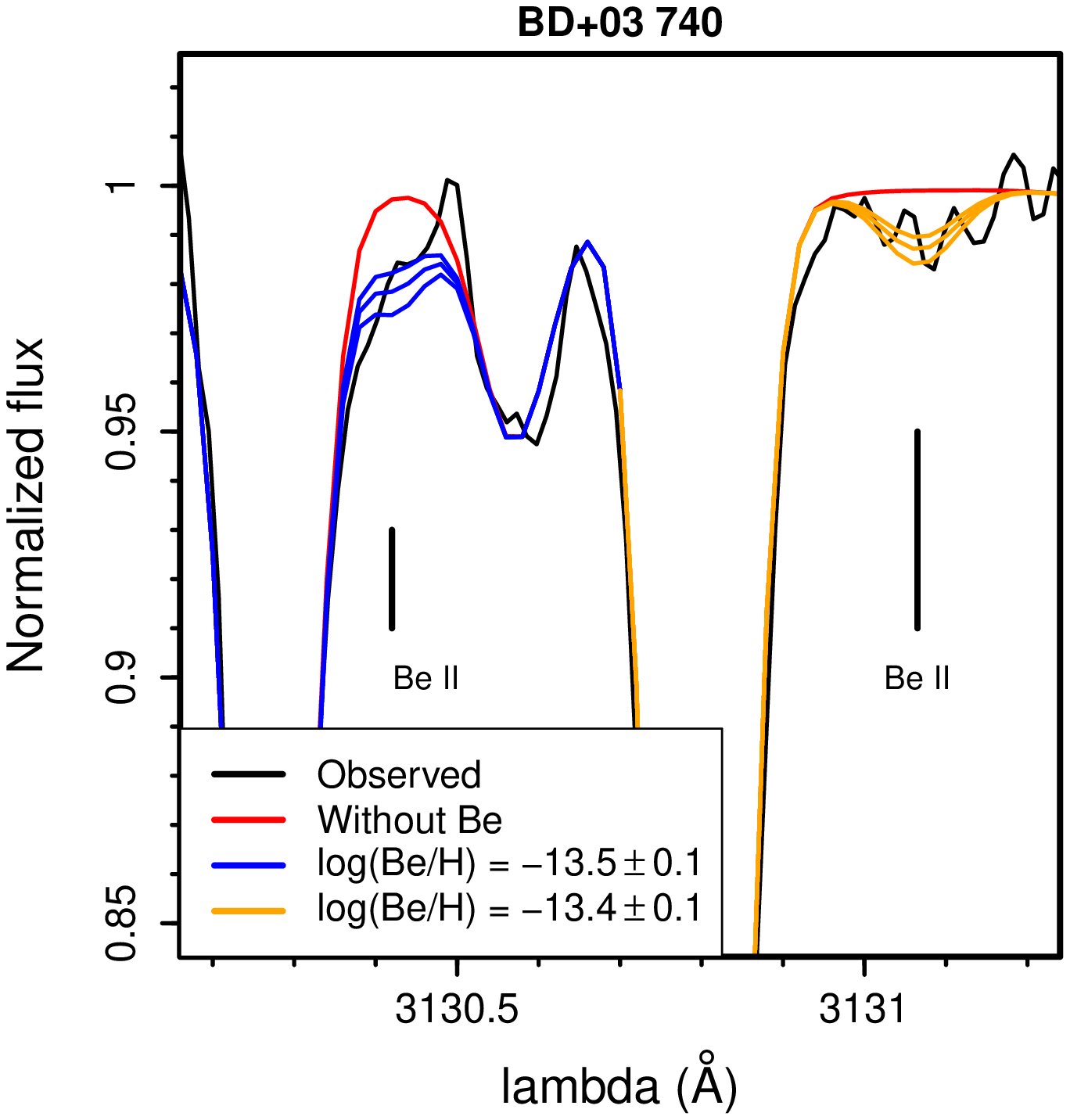}
        \includegraphics[height = 9cm]{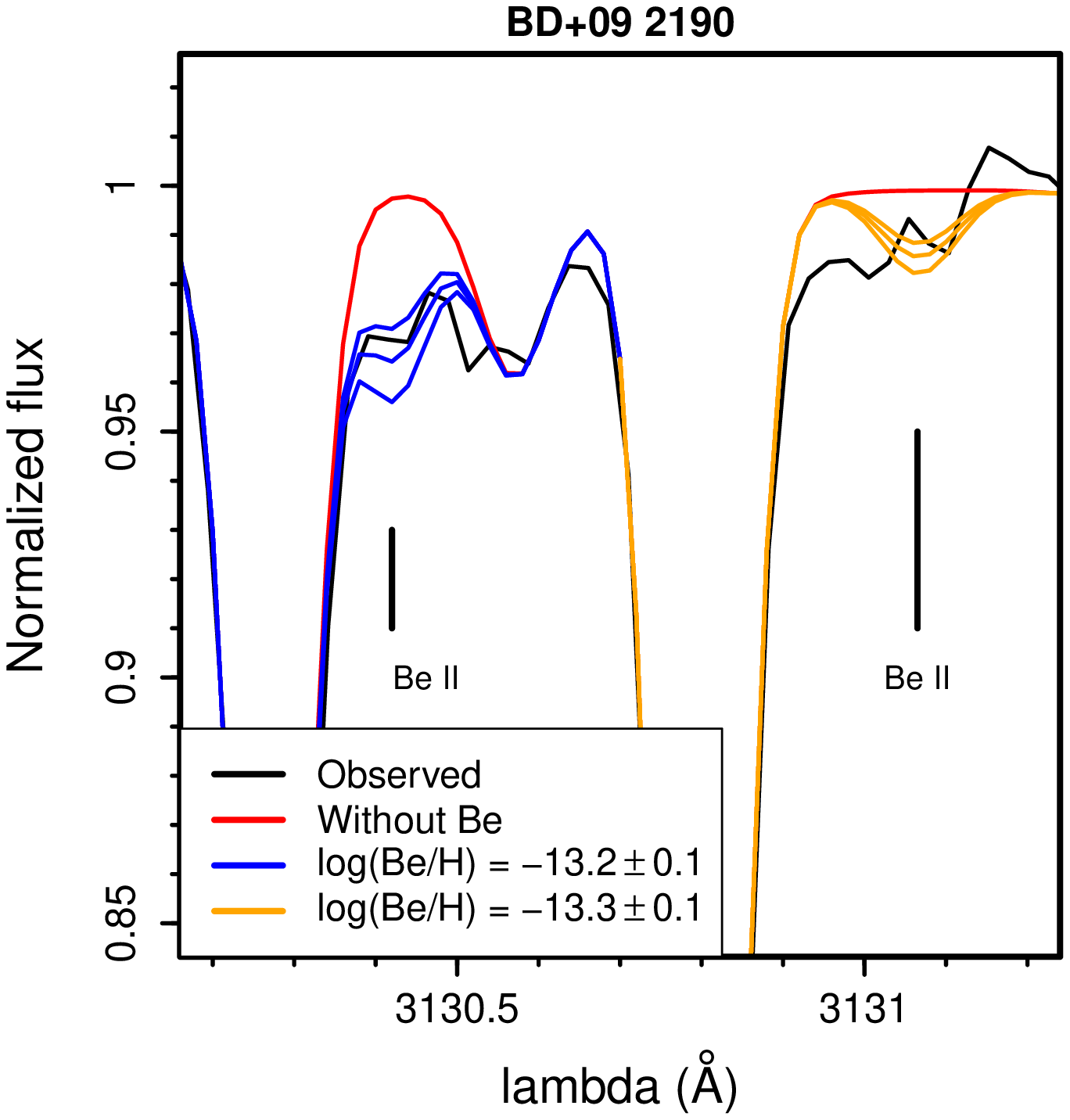}
        \caption{Synthetic fit to Be lines in stars BS 16968-061 (top left), G 64-12 (top right), BD+03 740 (bottom left), and BD+09 2190 (bottom right). Each of the Be lines was fit separately. The best fit is shown together with spectra where the Be abundance was changed by $\pm$0.1 dex. To aid in the visualisation, when necessary, some spectra are only shown in half of the panel.}\label{fig:be}%
\end{figure*}

In three cases, the two Be lines are fit with significantly different abundances (CD-33 1173, BD-13 3442, and BS 16968-061). In the cases of CD-33 and BS 16, one of the Be lines is not even properly detected. The reasons for these disagreements are unclear, but we have encountered such a situation before \citep{2009A&A...499..103S}, although in a somewhat more metal-rich regime. The disagreement does not seem to be caused by unknown blends, as we could not find common chemical peculiarities reported in the literature for these three stars. Moreover, we note that for G64-12 and G64-37, which have been reported to be carbon-enhanced metal-poor stars \citep{2016ApJ...829L..24P}, the two Be lines are consistently fit by essentially the same abundance value. BPS BS 16968-061, on the other hand, has a normal carbon abundance \citep{2013ApJ...762...26Y}.

A few abundances are marked in Table \ref{tab:atm} as having higher uncertainties because of a poor fit. Line 3131 of star BS 16968-061 is somewhat narrower than the line in the synthetic spectrum. We checked that there is no blending line affecting the profile of the synthetic line. To determine the best abundance, we restricted the fitting to the blue side of the observed line. Conversely, line 3131 observed in BD-13 3442 seems broader than the synthetic one. It also seems slightly shifted to bluer wavelengths. Line 3130 in G 64-12 seems narrower and shifted towards redder wavelengths than is indicated by the synthetic spectra. The profiles of line 3131 in the spectra of both G64-37 and BD+24 1676 seem strongly affected by noise. Even though the fitting procedure returns a significant detection, we opted to mark the fit as uncertain. As the Be lines in this metallicity regime are very weak, fitting the lines is a significant challenge even at the relatively high S/N of some of our spectra.

Since abundances from the two Be lines do not always agree, we decided to take a conservative approach guided by the spectrum synthesis. The calculations indicate that the line at 3130 \AA\ should always be the stronger of the doublet. Therefore, we are confident of a detection only when the 3130 line is detected. We estimated the uncertainties in the Be abundances, which come from the uncertainties in the atmospheric parameters and continuum placement, to be of the order of $\pm$ 0.15 dex. A few examples of fits to the Be lines are given in Fig. \ref{fig:be}. We  significantly zoomed over the Be lines to illustrate the difficulties in the fitting procedure.

We remark here that for star LP815-43 we were unable to detect the Be lines, even though we reanalysed the spectra from \citet{2000A&A...362..666P} where a detection was possible. The reason is that we were only partially successful in properly reducing the original data. Thus, the final co-added spectrum we had available was of lower a S/N than the one used in the original publication, preventing the detection of the Be lines.

\subsection{Comparison with previous results}

In this section, we present a comparison of our newly derived Be abundances with previous literature results. The comparison demonstrates the impact of using \emph{Gaia} DR2 parallaxes to assist on the computation of the surface gravity together with a temperature scale based on the IRFM. For stars CD-33 1173 and BPS BS 16968-061, we present Be abundances for the first time. Most stars in our sample have been analysed in multiple papers by Boesgaard and collaborators. In such cases, we compare our results only with the most recent of their papers \citep[i.e.][]{2011ApJ...743..140B}.

LP831-70: The beryllium abundance for this star was last determined by \citet{2011ApJ...743..140B}, and an upper limit of $\log(Be/H)$ $\leq$ $-$13.1 was found. Those authors determined $\log~g$ values using the ionisation equilibrium of Fe and Ti lines, finding $\log~g$ = 3.40. With our value of $\log~g$ = 4.42, we can provide more stringent limits to the Be abundance in this star; $\log(Be/H)$ $\leq$ $-$13.4 or $-$13.5.

BD+03 740: \citet{2011ApJ...743..140B} reported a detection at $\log(Be/H)$ = $-$13.4. Our abundance values are similar, $\log(Be/H)$ = $-$13.4 or $-$13.5. Our $\log~g$ value agree with theirs within 0.2 dex.

BD+24 1676: \citet{2011ApJ...743..140B} derived $\log(Be/H)$ = $-$13.28. Our values are considerably higher at $\log(Be/H)$ = $-$12.9 or $-$12.8. The difference in $\log~g$ values is of 0.3 dex. Our Be abundances are in better agreement with the higher metallicity of this star ([Fe/H] = -2.35).

BD+09 2190: Even though our $\log~g$ value differs from the one of \citet{2011ApJ...743..140B} by 0.3 dex, their Be abundance is the same as we found in this work: $\log(Be/H)$ = $-$13.22. The difference in $\log~g$ is balanced by a difference in 400 K in $T_{\rm eff}$, the value used in \citet{2011ApJ...743..140B} being cooler than ours. Our $T_{\rm eff}$ is the same as that obtained by \citet{2010A&A...512A..54C} with a different IRFM calibration. 

BD-13 3442: The beryllium abundance for this star was last determined by \citet{2011ApJ...743..140B} at a level of $log(Be/H)$ = $-$13.12. This is intermediate between the values we found for the two Be lines, $\log(Be/H)$ = $-$13.2 and $-$12.9. In our case, it does not seem possible to fit the two lines with the same Be abundance.

G64-12: The beryllium abundance for this star was first determined by \citet{2000A&A...364L..42P} at a level of $\log(Be/H)$ = $-$13.1. \citet{2011ApJ...743..140B} found an abundance of $\log(Be/H)$ = $-$13.43. Our atmospheric parameters, and hence also the Be abundance, are much closer to the values of \citet{2000A&A...364L..42P}. \citet{2011ApJ...743..140B} adopted $T_{\rm eff}$ = 6074 K and $\log~g$ = 3.72, in stark disagreement with the values we used.

\begin{figure*}
        \centering
        \includegraphics[height = 8.5cm]{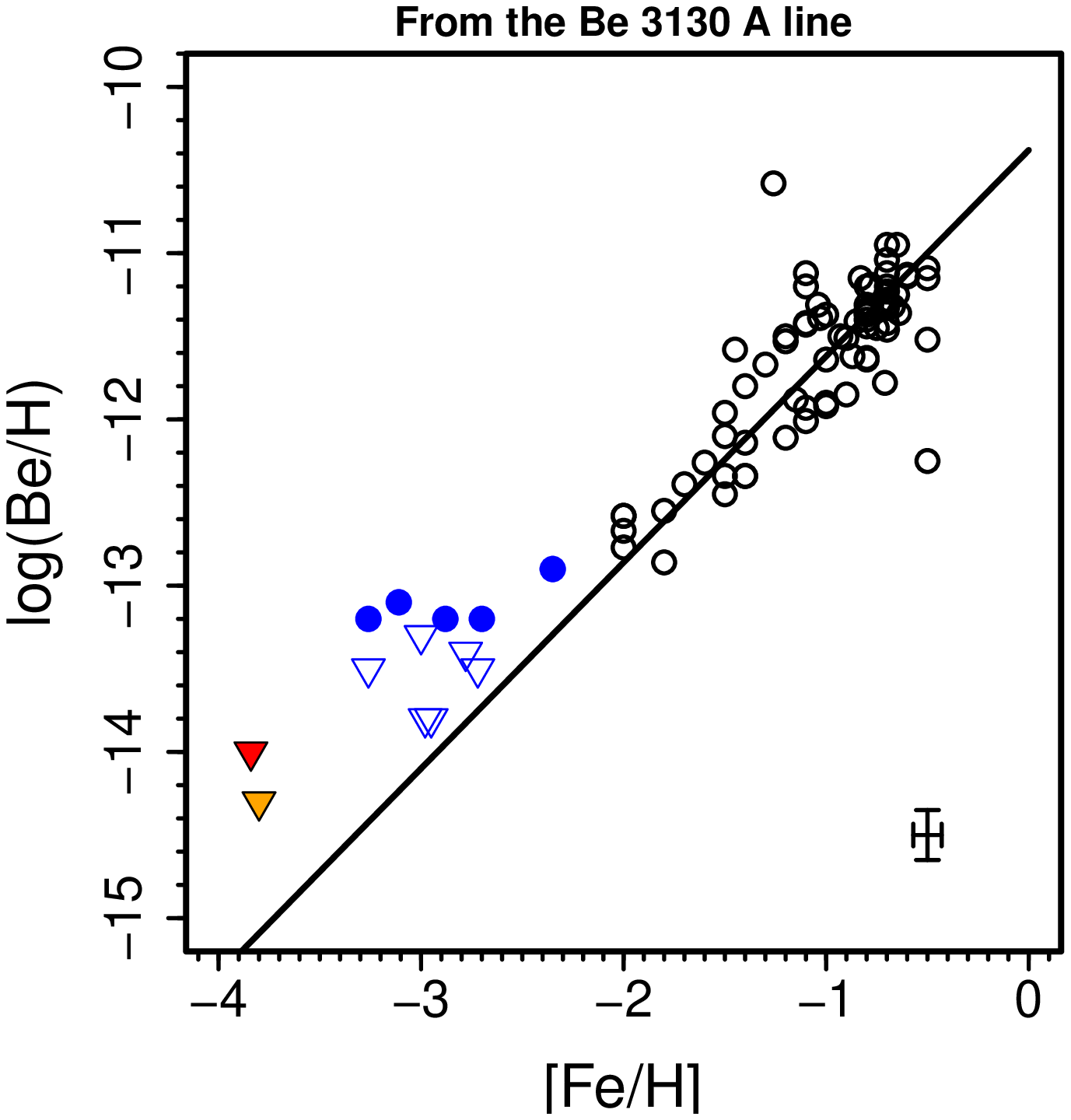}
        \includegraphics[height = 8.5cm]{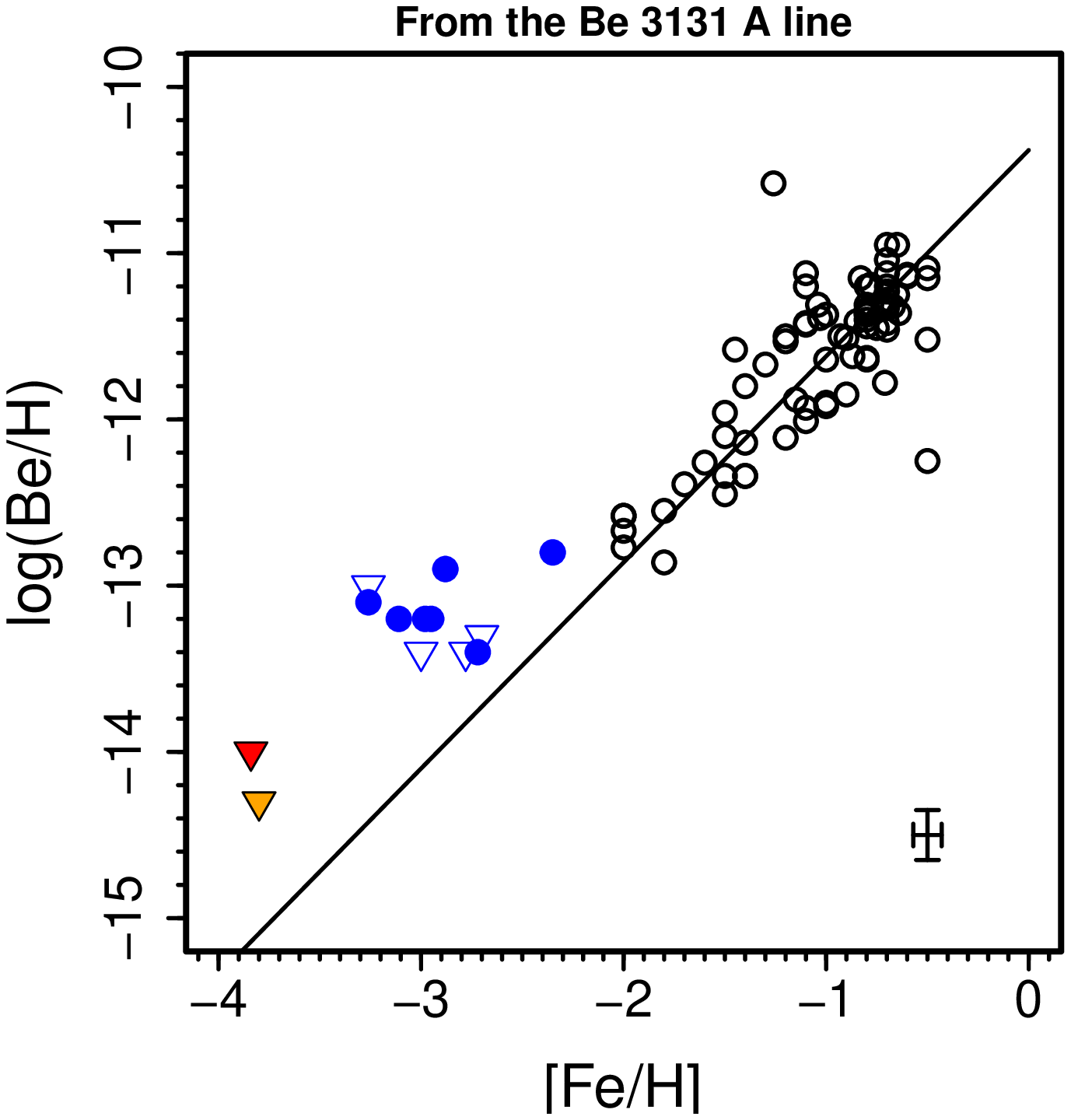}
        \caption{Beryllium abundances as a function of metallicity. Abundances resulting from this work are shown as blue circles and upper limits are shown as blue triangles. Stars from \citet{2009A&A...499..103S} are shown as open black circles. The solid line is a fit to those stars. The upper limits for stars 2MASS J18082002-5104378 \citep{2019A&A...624A..44S} and BD+44 493 \citep{2014ApJ...790...34P} are shown as upside-down red and orange triangles, respectively. The left and right panels show the abundances and limits obtained from the 3130 and 3131 \AA\ lines, respectively. A typical error bar of the new measurements is shown in the bottom right.}\label{fig:befeh}%
\end{figure*}

G64-37: \citet{2011ApJ...743..140B} found a value of $\log(Be/H)$ = $-$13.4. We derived higher values, $\log(Be/H)$ = $-$13.2 or $-$13.1. The comparison of atmospheric parameters is very similar to the case of G64-12. \citet{2011ApJ...743..140B} adopted $T_{\rm eff}$ = 6122 K and $\log~g$ = 3.87, which are in stark disagreement with respect to our values.

LP815-43: \citet{2000A&A...362..666P} obtained $\log(Be/H)$ = $-$13.1 for this star. \citet{2011ApJ...743..140B} reported $\log(Be/H)$ = $-$12.95. As mentioned before, we encountered several problems to reduce the UVES spectra of this star. Our upper limit is more stringent than previous results, but the confidence we place on it is not high.

CD-24 17504: \citet{2000A&A...362..666P} reported an upper limit of $\log(Be/H)$ $\leq$ $-$13.4 (analysing ESO VLT UVES spectra). On the other hand, \citet{2011ApJ...743..140B} reported a detection at $\log(Be/H)$ = $-$13.53 (analysing Keck HIRES spectra). Our analysis \citep[of the same spectra used in ][]{2000A&A...362..666P} can only put upper limits and at different levels for each Be line. Within those limitations, our results agree with the two literature sources above.

\section{Discussion}\label{sec:discussion}

\subsection{Flattening of the relation between Be and metallicity}

The beryllium abundances obtained in this work are shown in diagrams of $\log(Be/H)$ as a function of [Fe/H] in Fig.\ \ref{fig:befeh}. The few detected abundances for the extremely metal-poor stars clearly deviate from the linear relation defined by the stars of higher metallicity analysed by \citet{2009A&A...499..103S}. 

For illustration, in Fig.\ \ref{fig:befeh} we display separated plots for the results obtained from the two Be lines. The deviation from the linear relation is very striking with the results from the 3131 \AA\ line (right panel). Nevertheless, the results from the 3130 \AA\ line (left panel) also clearly show that the same linear relation with [Fe/H] is not valid at this regime. 

We note here that this change in behaviour has been noticed before. \citet{2009ApJ...701.1519R}, for example, suggested that two linear relations, with different slopes, are needed to fit the data. These authors speculated that this behaviour could indicate a change in the main mechanism contributing to the Be production. At the lowest metallicities, the contribution from accelerated C, N, and O atoms would dominate. Above [Fe/H] $\sim$ $-$2.0, cosmic rays destroying C, N, and O atoms of the interstellar medium would dominate the Be production. However, theoretical models indicate that the first mechanism always dominates, even up to the highest metallicities where the second production channel is responsible for at most 25-40\% of Be  \citep{2012A&A...542A..67P}. 

Four stars with metallicities ranging from [Fe/H] = $-$3.26 to $-$2.70 dex have, within the uncertainties, essentially the same Be abundance around $-$13.2 dex. The two stars that deviate the most from the linear fit (G64-12 and G64-37) have Be abundances more than 7 or 8$\sigma$ above what would be expected for stars of their metallicities. Given the uncertainties, there is no easy way to reconcile the measurements with the linear relation.

In the extremely metal-poor stars analysed here, the detected abundances create an upper envelope of points that indicates a flattening of the Be-versus-Fe relation. However, the number of upper limits from this work, together with the upper limits from the literature for two stars of lower metallicity, suggest that this flattening is not a universal feature. It suggests instead that we are looking at an increased scatter in the distribution of abundances.

Moreover, we note that theoretical models of Be production through cosmic-ray nucleosynthesis, such as the ones of \citet{2012A&A...542A..67P}, obtain a linear relation between Be and Fe down to the lowest metallicities. The fact that the observations do not confirm these predictions suggests that the expectations for Be production in the early Galaxy might require some revision. Of course, very little is known about the cosmic rays, and consequently about cosmic-ray-induced nucleosynthesis, in the early Galaxy.

\subsection{The Gaia Enceladus merger event}

\begin{figure}
        \centering
        \includegraphics[height = 8.5cm]{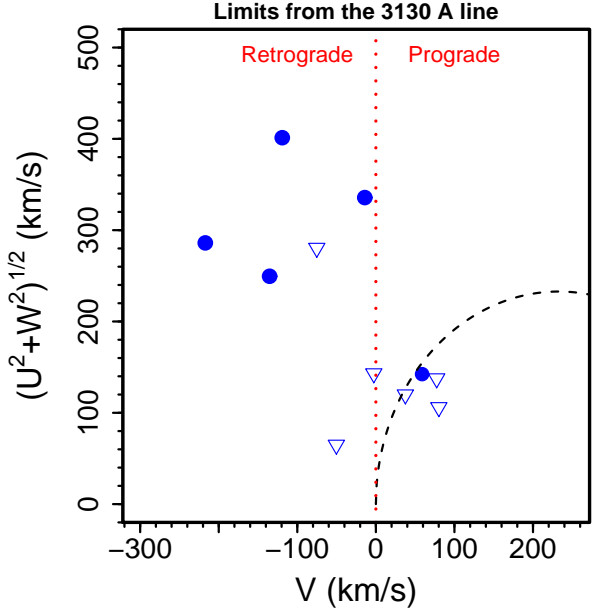}        
        \caption{Toomre diagram for the sample of extremely metal-poor stars. Symbols are as in the left panel of Fig.\ \ref{fig:befeh}. Error bars (typically below 5 km s$^{-1}$) are smaller or equivalent to the size of the symbols. The black dashed line indicates a total space velocity of 232 km s$^{-1}$.}\label{fig:toomre}%
\end{figure} 

\citet{2005A&A...436L..57P} were the first to show that different Galactic kinematic stellar populations \citep[the so-called accretion and dissipative components, see][]{2003A&A...406..131G} could be separated in a diagram of [O/Fe] as a function of $\log(Be/H)$. Afterwards, \citet{2009A&A...499..103S}, in their analysis of Be abundances in metal-poor halo and thick-disc stars, found that the halo stars are the ones to divide into two different sequences in a plot of [$\alpha$/Fe] as a function of $\log(Be/H)$. They suggested the division was a sign of either an accreted stellar component or of variations in the star formation history in initially independent regions of the early halo.

Recently, thanks to \emph{Gaia} data \citep{2018A&A...616A..10G,2018A&A...616A..11G}, it became apparent that the Milky Way suffered a major merger about 8-10 Gyr ago. The merging object has been called Gaia Enceladus \citep{2018Natur.563...85H} or Gaia Sausage \citep{2018MNRAS.478..611B}. The total mass of the dwarf-galaxy progenitor could be of the order of 10$^{8.85-9.85}$ M$_{\odot}$ \citep{2020MNRAS.497..109F}. 

\citet{2020MNRAS.496.2902M} used literature data to compare the evolution of Be abundances in stars thought to belong to Gaia Enceladus with those of stars formed in-situ. They found the behaviour of the two populations, in the $\log(Be/H)$ versus [Fe/H] plot, to be similar at low metallicity, but to deviate with increasing [Fe/H]. \citet{2020MNRAS.496.2902M} argued that the observed scatter in a diagram of $\log(Be/H)$ versus [Fe/H] could thus be related to different stellar components mixed inside the Galactic halo. 

We used the kinematic and orbital data of our sample stars (Tables \ref{tab:kinem} and \ref{tab:orbits}) to investigate whether we can identify stars from the Gaia Enceladus merger. Figure \ref{fig:toomre} displays the Toomre diagram of the stars. The velocities in the direction of Galactic rotation are mostly small or negative. This already indicates some probability of finding Gaia Enceladus stars, which seem to have preferentially retrograde motions \citep{2018Natur.563...85H}.

\begin{figure}
        \centering
        \includegraphics[height = 8.5cm]{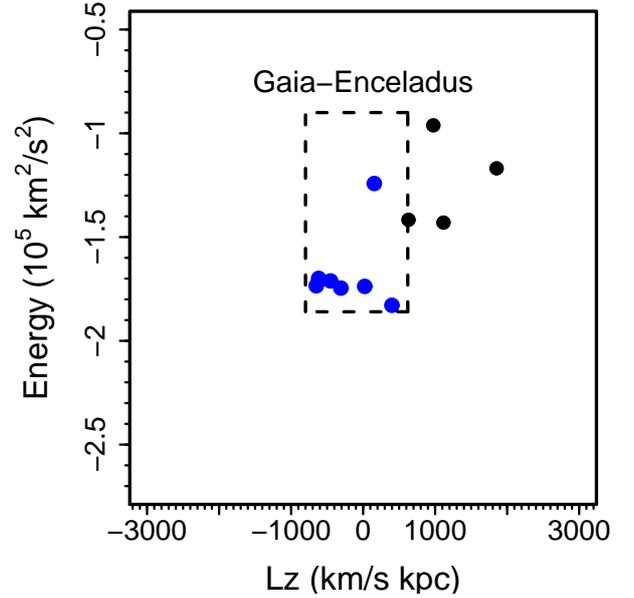}    
        \caption{Diagram of energy as a function of the z-component of the angular momentum. The box defining the Gaia Enceladus region follows \citet{2019A&A...630L...4M}. Stars inside this region are marked in blue, while those outside are shown in black.}\label{fig:ener-lz}%
\end{figure}

Figure \ref{fig:ener-lz} displays the stars in a diagram of energy as a function of the orbital angular momentum (the z-component of the angular momentum, L$_z$). To separate the stars belonging to Gaia Enceladus, we adopted the criteria from \citet{2019A&A...630L...4M} that makes use of orbital energy, of L$_z$, and of the component of the angular momentum perpendicular to L$_z$ (L$_{perp}$). The L$_z$ component is given by the {\sf GalPot} orbit integration (Section \ref{sec:data}). The L$_{perp}$ component, on the other hand, we computed directly using the current positions and velocities of the stars. 

Three stars are clearly outside the energy-L$_z$ box that defines Gaia Enceladus (BD+09 2190, G64-12, G64-37). One star, just at the edge, still agrees with the Gaia Enceladus definition within the uncertainties, even though formally it falls outside the box (CD$-$24 17504). All other stars (which are most of the sample) are consistent with having originated at the Gaia Enceladus progenitor. We note here that \citet{2020MNRAS.496.2902M} seem to consider G64-37 part of Gaia Enceladus, even though in our Fig.\ \ref{fig:ener-lz} the star clearly falls outside the relevant region. Only two of our other stars were included in the \citeauthor{2020MNRAS.496.2902M} sample (BD+24 1676 and BD+09 2190), and in these two cases our analyses agree that they likely originated in the Gaia Enceladus progenitor.

\begin{figure}
        \centering
        \includegraphics[height = 8.5cm]{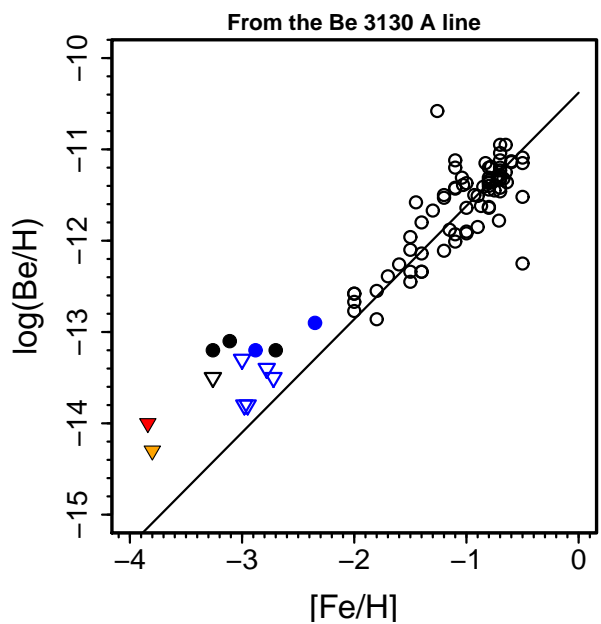}
        \caption{Beryllium abundances as a function of metallicity. The points are the same as in the left panel of Fig.\ \ref{fig:befeh}, except that we colour the four extremely metal-poor stars found to likely be in-situ halo stars in Fig.\ \ref{fig:ener-lz} in black.}\label{fig:befeh.ge}%
\end{figure}

Finally, in Fig.\ \ref{fig:befeh.ge} we again show the diagram of $\log(Be/H)$ as a function of metallicity, but separating the extremely metal-poor stars likely belonging to Gaia Enceladus (in blue) from those that were likely formed in-situ at the Galactic halo (in black). Interestingly, it seems here that the stars from Gaia Enceladus are the ones closer to the extension of the linear relation, while those that deviate the most are in-situ stars.

In particular, we should remark here the position of stars BPS BS 16968 and CD-33 1173, for which we attempted to derive Be abundances for the first time. These stars are the two open blue triangles plotted over each other at about [Fe/H] $\sim$ $-$3.0 and $\log(Be/H)$ $\sim$ $-$13.8 in Fig.\ \ref{fig:befeh.ge}. They are the most important points that demonstrate how stars from Gaia-Enceladus remain closer to the linear relation in a wide range of metallicities. 

We believe any extended discussion about Be abundances in stellar sub-components of this regime is for now precluded by our small sample. Nevertheless, two main results can be obtained from our analysis: i) the deviation from the linear relation at [Fe/H] $\sim$ $-$3.0 is real, although not universal; and ii) the scatter we observed at these metallicities seems related to the inhomogeneous nature of the early halo.%

\subsection{Inhomogeneous star formation in the early Galaxy}

Having established our observational results, we now searched for a possible explanation to the increased scatter of points in the $\log(Be/H)$ versus [Fe/H] at the extremely metal-poor regime. To do that, we summarised results from an extensive body of investigations that have looked at the details of the first phases of Galactic chemical enrichment.

The interpretation we would like to advance is that the main drivers of the observed scatter are the abundances of Fe and not those of Be. Or, in other words, that at the time of the formation of the early halo, Fe abundances (in a Galactic context) were more inhomogeneous than the Be abundances.

\subsubsection{Early enrichment in beryllium}

The idea that abundances of Be (and of other cosmic-ray spallation products) can have smaller scatter when compared to abundances of elements produced by stellar nucleosynthesis was first suggested by the results of \citet{1999ApJ...522L.125S} and \citet{2001ApJ...549..303S}. These authors computed chemical evolution models of an inhomogeneous early Galactic halo. Based on these models, Be abundances were suggested to be a good cosmochronometer for the early stages of the Galaxy \citep{2000IAUS..198..425B,2001ApJ...549..303S}. This suggestion was tested by \citet{2004A&A...426..651P,2007A&A...464..601P} who derived Be abundances in turn-off stars of the globular clusters NGC 6397 and NGC 6752. These measurements were very challenging, but ages obtained comparing the Be abundances with predictions of chemical evolution models \citep[from][]{2002ApJ...566..252V} were shown to be in excellent agreement with ages derived using isochrones.

\citet{2001ApJ...549..303S} seem to have made the last attempt to follow the Be nucleosynthesis in a model of inhomogeneous Galactic chemical enrichment. In their models, at an age of 0.2 Gyr (when the mean [Fe/H] is close to $-$3.0), the Galaxy has a dispersion of about 1.2 dex in Fe and 0.8 dex in Be. At 0.4 Gyr (when the mean [Fe/H] is already close to $-$2.3), the dispersion is of about 0.7 dex in Fe and 0.4 dex in Be. More recent theoretical studies of cosmic-ray nucleosynthesis, such as that of \citet{2012A&A...542A..67P}, make use of the so-called instantaneous mixing approximation, meaning it is assumed that at any moment the Galactic material is well mixed. Such models cannot make predictions about the abundance dispersion at a given moment. We hope our new results can trigger the development of new models that investigate the inhomogeneous enrichment of Fe and Be in the early Galaxy.

An alternative model of Be production in the early Galaxy was presented by \citet{2008ApJ...673..676R}. These authors explored the possibility of cosmic-ray nucleosynthesis in the intergalactic medium at high redshift. Potentially, an abundance level of about $\log(Be/H)$ $\sim$ $-$13 could be reached at the time of the Galaxy formation (redshift $\sim$ 3). If correct, this would nicely explain the flattening trend we observe in our abundances. Nevertheless, the same model overproduces $^{6}$Li, and it is thus not clear how applicable it could really be.

\subsubsection{Early enrichment in iron}

The inhomogeneous enrichment of the early Galactic medium in stellar nucleosynthetic products has been the subject of several studies. \citet{1995ApJ...451L..49A} seem to have been the first to suggest that a high dispersion in metallicity should be present in the early Galaxy, if the medium was enriched by only few supernova (SN) events characterised by a discrete range of progenitor masses. Several models of inhomogeneous chemical evolution of the early Galaxy have shown that an abundance scatter can be present for stars with metallicities below about $-$3.0 \citep{2000A&A...356..873A,2003MNRAS.339..849O,2005A&A...436..879K}. In the work of \citet{2000A&A...356..873A}, at 0.2 Gyr the total range of possible [Fe/H] values is as wide as about 2 dex. Nevertheless, other works have suggested that a degree of chemical homogeneity might still be present at the level of stars formed inside the same stellar cluster \citep{2010ApJ...721..582B}.

Additional evidence of an inhomogeneous early metal enrichment comes from cosmological simulations. Some of these simulations try to follow the metallicity of the material that is forming stars inside dark matter mini haloes. In particular, there are several works that have tried to study the conditions for the formation of second-generation stars from material that has been contaminated by the ejecta of the first SNe of Population III (Pop III).

The interstellar medium metallicities produced by mixing one single enrichment event (i.e. material produced by one single Pop III SN) with primordial gas can be very heterogeneous. It could create a total range from [Fe/H] = $-$5.0 to $-$3.0 or higher, which would result in second-generation stars formed with very different metallicities \citep{2014MNRAS.444.3288J,2015MNRAS.452.2822S,2017ApJ...844..111C}.
 
One possible reason for such large early scatter is anisotropy in the ejecta, resulting in dissimilar amounts of the same element being dispersed in different directions. A second reason is that distinct parts of the ejecta can have varied values of entropy, and they each cool on specific timescales, effectively contributing differently to the material that will form a new generation of stars \citep{2015MNRAS.451.1190R,2016MNRAS.456.1410S}.

Evidently, we are not arguing that all our sample stars have been directly produced as second-generation stars after one Pop III SN.\footnote{We mention here that \citet{2020A&A...643A..49H} recently suggested that one of our sample stars, BD+09 2190 with [Fe/H] = $-$2.7, could actually be one such second-generation star. This observation seems to agree with our suggested interpretation of the Be and Fe abundance behaviour at the extremely metal-poor regime. It additionally indicates that the level of $\log~(Be/H)$ $\sim$ $-$13.2 was reached very early in the formation of the Galaxy.} The important consideration is that a large body of literature results demonstrate the plausibility of assuming a large scatter in Fe abundances for stars being formed in the early Galaxy. It seems that when investigating the extremely metal-poor regime, we are starting to probe a different phase of star formation; one where mixing is inefficient and the medium has a high degree of inhomogeneity. This is the same reasoning used to explain the increased scatter observed in the abundance of r-process elements in stars below [Fe/H] $\sim$ $-$3.0 \citep{2007A&A...476..935F,2019arXiv190101410C}. Our argument is that current evidence supports the idea that it is Fe, and not Be, that is the main driver of the scatter we observe at the extremely metal-poor regime.

A similar possibility of increased scatter at low metallicities was discussed by \citet{2019A&A...624A..44S}, but mainly relied on the relation between $\log(Be/H)$ and [O/H]. Unfortunately, we cannot derive reliable oxygen abundances here, as they require the use of three-dimensional hydrodynamical models. The fact that it is Fe (or O) driving the observed scatter does not necessarily support the idea of Be as a cosmochronometer, but this indicates that Be abundances might indeed be very important as tracers of early star forming conditions in the Galaxy.

\section{Conclusions}\label{sec:end}

In this work, we revisited the abundances of Be in a sample of nine extremely metal-poor stars that have been observed with the UVES spectrograph at the VLT. In addition, we presented an analysis of the Be lines in stars BPS BS 16968-0061 and CD-33 1173 for the first time. The new near-UV spectra of BPS BS 16968-0061 that were analysed here required a total of 20h of observations. Atmospheric parameters, and surface gravity in particular, were recomputed taking advantage of \emph{Gaia} parallaxes and photometric data. Determining reliable Be abundances for these stars is very challenging, because of the weakness of the Be lines. In many cases, we could only place upper limits or derive uncertain abundance values.

Two main conclusions could be drawn from our new analysis of Be in extremely metal-poor stars. First, the few measurements allow us to confirm an increased scatter of the points in a $\log(Be/H)$ -versus-[Fe/H] plot at this metallicity regime. A clear deviation from the linear relation defined at higher metallicities is seen. Stars with metallicities changing by more than 0.5 dex have very similar Be abundances of the order of $\log(Be/H)$ = $-$13.2 dex. Nevertheless, a number of upper limits (from this work and from the literature) show that there is no Be abundance plateau.

Moreover, we identify seven stars that seem to belong to the Gaia Enceladus progenitor (the association of an eighth star being more uncertain). This leads to our second conclusion, which is that the co-existence of in-situ and accreted stars is likely behind the observed scatter. Stars from Gaia Enceladus seem to follow the linear relation between Fe and Be to low metallicities. Stars formed in situ seem to deviate from this relation. The two stars that are newly analysed here have been important to demonstrate the behaviour of the Be abundances of stars from Gaia Enceladus in a wider range of metallicities.

To explain our observations, we suggest that it is Fe, and not Be, that is the main driver of this increased scatter. Cosmological simulations of the formation of second-generation stars and inhomogeneous chemical evolution models seem to support a large scatter of stellar nucleosynthetic products very early in the Galaxy. For Be, on the other hand, there is evidence that its cosmic-ray origin might result in a more homogeneous abundance distribution across the Galaxy. It is therefore the Be abundances, and not the Fe ones, that most likely offer the best clock to understanding early Galactic chemical enrichment. In that sense, what we uncovered in this work is observational evidence for the inhomogeneous Fe enrichment at a given time.

In this context, Be abundances in extremely metal-poor stars might become a very important tracer of early star forming conditions in the Galaxy. It is clearly necessary to extend the sample of such stars with determined Be abundances. Such observations are, however, very challenging. Essentially, all well known bright extremely metal-poor stars have already been investigated for Be. It would still be important to re-observe these stars, in an attempt to convert some of the upper limits to detections. However, unless new, previously unknown, relatively bright, extremely metal-poor stars are identified, the best chance for increasing the sample seems to be in next generation instrumentation. The advent of the Cassegrain U-Band Efficient Spectrograph (CUBES),\footnote{\url{http://cubes.inaf.it/}} a new near-UV spectrograph planned for the VLT, will open important parameter space in this sense \citep{2014Ap&SS.354...55S,2018SPIE10702E..2EE}. It is expected that CUBES will be able to obtain spectra for stars 2-3 mag fainter than what is currently possible with UVES. This will potentially make a sample of hundreds of extremely metal-poor stars available for Be abundance determination.

\begin{acknowledgements}
We thank the anonymous referee for the useful comments that improved the manuscript. R.S. acknowledges support by the Polish National Science Centre through project 2018/31/B/ST9/01469. M.G.Z. thanks the Nicolaus Copernicus Astronomical Center for the summer research internship position that enabled his participation in this work. This research has made use of: NASA's Astrophysics Data System; the SIMBAD database, operated at CDS, Strasbourg, France; the VizieR catalogue access tool, CDS, Strasbourg, France. The original description of the VizieR service was published in \citet{2000A&AS..143...23O}; data products from the Two Micron All Sky Survey, which is a joint project of the University of Massachusetts and the Infrared Processing and Analysis Center/California Institute of Technology, funded by the National Aeronautics and Space Administration and the National Science Foundation; the VALD database, operated at Uppsala University, the Institute of Astronomy RAS in Moscow, and the University of Vienna.. The analysis has made extensive use of {\sf R} \citep{Rcore}, {\sf RStudio} \citep{RStudio}, and the {\sf R} packages {\sf gplots} \citep{gplots}, and {\sf pracma} \citep{pracma}. This work has made use of data from the European Space Agency (ESA) mission {\it Gaia} (\url{https://www.cosmos.esa.int/gaia}), processed by the {\it Gaia} Data Processing and Analysis Consortium (DPAC, \url{https://www.cosmos.esa.int/web/gaia/dpac/consortium}). Funding for the DPAC has been provided by national institutions, in particular the institutions participating in the {\it Gaia} Multilateral Agreement.
\end{acknowledgements}

\bibliographystyle{aa} 
\bibliography{../../../Biblio/smiljanic} 

\end{document}